\documentclass[final,3p,times]{elsarticle}

\usepackage{amsmath}
\usepackage{bbm}
\usepackage{bm}
\usepackage{mathrsfs}
\biboptions{numbers,sort&compress}

\newcommand\imag{i}                  
\newcommand\vek[1]{\bm{#1}}          
\newcommand\gr[1]{\mathrm{#1}}       
\newcommand{\de}{\partial}           
\DeclareMathOperator{\La}{\mathscr L}
\newcommand\eps{\varepsilon}
\newcommand\openone{\mathbbm{1}}
\newcommand\vp{\varphi}

\journal{Nuclear Physics A}

\newcommand{\cor}[2]{{}{#2}} 

\begin{document}

\begin{frontmatter}

\title{Spin-one color superconductors: collective modes and effective
Lagrangian}
\author[ustc]{Jin-yi Pang}
\ead{pangjin@mail.ustc.edu.cn}
\author[rez]{Tom\'a\v{s} Brauner\fnref{ascr}}
\ead{tbrauner@physik.uni-bielefeld.de}
\fntext[ascr]{On leave from Department of Theoretical Physics, Nuclear Physics
Institute ASCR, CZ-25068 \v{R}e\v{z}, Czech Republic}
\author[ustc]{Qun Wang}
\ead{qunwang@ustc.edu.cn}
\address[ustc]{Interdisciplinary Center for Theoretical Study and
Department of Modern Physics, University of Science and Technology of China,
Anhui 230026, People's Republic of China}
\address[rez]{Faculty of Physics, University of Bielefeld, D-33501 Bielefeld,
Germany}

\begin{abstract}
We investigate the collective excitations in spin-one color superconductors.
We classify the Nambu--Goldstone modes by the pattern of spontaneous symmetry
breaking, and then use the Ginzburg--Landau theory to derive their dispersion
relations. These soft modes play an important role for the low-energy dynamics
of the system such as the transport phenomena and hence are relevant for
late-stage evolution of neutron stars. In the case of the color-spin-locking
phase, we use a functional technique to obtain the low-energy effective action
for the physical Nambu--Goldstone bosons that survive after gauging the
color symmetry.
\end{abstract}


\end{frontmatter}


\section{Introduction}
\label{Sec:Intro}

In the phase diagram of quantum chromodynamics (QCD), cold quark
matter is expected to deconfine at high density and be in a color
superconducting state (see \cite{Alford:2007xm,Wang:2009xf,
Huang:2010nn,Fukushima:2010bq} for recent reviews). Understanding
the properties of matter under such extreme conditions is, apart
from being an integral part of the quest for the fundamental laws of
nature, relevant for the astrophysics of compact stellar objects.
The density in their cores is presumably high enough to form
deconfined quark matter.

Since the energy scale for strong interaction in astrophysical
processes mentioned above is of order $100\,\text{MeV}$, one can
safely neglect the heavy quark flavors and only consider the three
lightest ones, up ($u$), down ($d$), and strange ($s$). At densities
so high that the masses of all light quarks can be neglected, the
ground state of three-flavor quark matter is known to be the
color-flavor-locking (CFL) state. However, as the chemical potential
drops to the range interesting for astrophysical applications,
the strange quark mass is not negligible and starts to play an
important role. It reduces the Fermi sea of $s$ quarks, leading to an excess
electric charge which must in turn be compensated by an imbalance between the
$u$ and $d$ quarks. As a consequence, the highly symmetric CFL state feels
stress and can give way to other pairing patterns \cite{Rajagopal:2005dg}.

If the stress on the CFL pairing induced by the strange quark mass
is too high only the $u$ and $d$ quarks pair, which is usually
denoted as the 2SC phase. However, as explained above, the energy
gain from the 2SC pairing is reduced by the requirement of electric
charge neutrality. It may then happen that the mismatch between the
Fermi levels is so large that pairing between quarks of different
flavors is completely ruled out. In such a case, one is still left
with the possibility of pairing quarks of the \emph{same} flavor.
Nevertheless, since the QCD-induced effective quark--quark
interaction is attractive in the color-antisymmetric channel, this
requires the wave function of the Cooper pair to be symmetric in
Lorentz indices by the Pauli principle, i.e. to carry nonzero spin.
While spin one is the simplest possibility, the ground state will in
general be a mixture of all partial waves with an odd spin
\cite{Feng:2007bg}.

There are several situations in which spin-one pairing may occur.
Apart from the pairing of quarks of a single flavor considered
originally in
Refs.~\cite{Bailin:1983bm,Schaefer:2000tw,Hosek:2000fn,Alford:2002rz},
there is another possibility of pairing of quarks of the same
\emph{color} \cite{Buballa:2002wy}. This naturally appears in the
2SC phase where only quarks of two colors are involved in the
pairing and the third one is left over. We will use this proposal as
a warm-up exercise in Sec.~\ref{Sec:1color}. However, in the rest of
the paper, we will focus on the more likely pattern of pairing of
quarks of the same flavor. Even then one can distinguish different
scenarios. When cross-flavor pairing is completely prohibited, all
three flavors will pair in some of the spin-one states. On the other
hand, only $s$ quarks may undergo spin-one pairing as a complement
to the 2SC state. We will have in mind mostly the first case, since
in the latter the presence of the color-asymmetric 2SC state can
modify the ground state of the spin-one phase \cite{Alford:2005yy}.

The classification of the phases of a single-flavor spin-one color
superconductor was worked out in Refs.~\cite{Schmitt:2004et,Brauner:2008ma}. The
physical properties of the so-called inert spin-one phases were investigated in
Refs.~\cite{Schmitt:2002sc,Schmitt:2003xq,Schmitt:2003aa}. Astrophysical
implications of the presence of a spin-one phase were discussed in
Refs.~\cite{Schmitt:2005ee,Schmitt:2005wg,%
Aguilera:2005tg,Aguilera:2006cj,Aguilera:2006xv,Sa'd:2006qv,Wang:2006tg,
Blaschke:2008br}.
An alternative approach to spin-one color superconductors, based on
the Schwinger--Dyson equations, was taken in Ref.~\cite{Marhauser:2006hy}. For
some recent references on the topic see
Refs.~\cite{Feng:2009vt,Wang:2009if,Wang:2010yd}.

It was shown in Ref.~\cite{Brauner:2008ma} that as a consequence of
weak interactions, the true ground state of spin-one color
superconductors is an inhomogeneous state with helical ordering,
similar to what happens in non-center-symmetric ferromagnets
\cite{Dzyaloshinsky:1958dz,Moriya:1960mo}. In the present paper this
additional structure will not, for the sake of simplicity, be
considered. This may be understood as restricting to length scales
much larger than the average distance between quarks, yet much
smaller than the wavelength of the helix.

The plan of the paper is as follows. In Sec.~\ref{Sec:1color} we
shall investigate the collective modes in a single-color spin-one
superconductor \cite{Buballa:2002wy}. This elucidates some of the
peculiarities of spontaneous breaking of spacetime symmetry
\cite{Low:2001bw} without the technical complications brought by a
complex matrix order parameter. In particular, we will show an
explicit example of a phase in which the number of Nambu--Goldstone
(NG) bosons is smaller than the number of broken symmetry generators
\cite{Nielsen:1975hm} (see Ref.~\cite{Brauner:2010wm} for a recent
review including more references). Also, it will be demonstrated
that due to the fact that the order parameter breaks rotational
symmetry, the division of the NG modes into multiplets of the
unbroken symmetry holds only in the long-wavelength limit. Nonzero
momentum of the mode breaks the remaining symmetry and results in
further splitting of the multiplets.

Section \ref{Sec:1flavor} is the main content of the paper. We
analyze the spectrum of collective modes in a spin-one color
superconductor, elaborating on the classification of the NG modes
suggested in our previous paper \cite{Brauner:2009df}. For
simplicity, we consider quark matter composed of one quark flavor
only. This assumption is released in Sec. \ref{Sec:CSL_EFT} where we
concentrate on the color-spin-locking (CSL) phase in electrically
neutral three-flavor quark matter. Using a generalization of a trick
due to Son \cite{Son:2002zn}, we construct the low-energy effective
Lagrangian for the physical NG bosons that remain in the spectrum
after gauging the color and electromagnetic sectors of the symmetry
group. Finally, in Sec. \ref{Sec:conclusions} we summarize and make
conclusions.


\section{Single-color spin-one superconductor}
\label{Sec:1color}

The pairing pattern considered here was suggested in Ref.~\cite{Buballa:2002wy}.
Since it complements the standard 2SC state,
it involves quarks of the two lightest flavors and one color. They
pair in a flavor-singlet state, and the wave function must therefore
be symmetric in the Lorentz indices. In the simplest case of spin
one, the order parameter is a flavor singlet and spin triplet, and
thus is represented by a complex vector of the $\gr{SO(3)}$
rotational group.

We will analyze the excitation spectrum using a Lagrangian
which can be regarded as the time-dependent
Ginz\-burg--Landau (GL) theory. We will demand that the Lagrangian has
rotational as well as $\gr{U(1)}$ phase invariance, corresponding to
the conservation of particle number, \cor{}{and preserves parity}. 
The most general Lagrangian for a complex vector field $\vek\phi$ that has the required symmetries
and includes operators of canonical dimension up to four, reads
\begin{equation}
\La=ic_1{\vek\phi}^\dagger\cdot\de_0\vek\phi+
c_2\de_0{\vek\phi}^\dagger\cdot\de_0\vek\phi
-a_1\de_i{\vek\phi}^\dagger\cdot\de_i\vek\phi
-a_2|\vek\de\cdot\vek\phi|^2 - b{\vek\phi}^\dagger\cdot\vek\phi
-d_1({\vek\phi}^\dagger\cdot\vek\phi)^2 - d_2|\vek\phi\cdot\vek\phi|^2,
\label{Lagrangian_1color}
\end{equation}
where the spatial vectors are in boldface and their inner product is
indicated by a dot; we will use this convention throughout the
paper. \cor{}{One should keep in mind that by rescaling the field appropriately, one can get rid of
one of the unknown coefficients, say, $c_1$.} The static part of this Lagrangian was investigated in
\cite{Buballa:2002wy}, so we just summarize the results here.
First of all, boundedness of the potential from below demands that
$d_1>0$ and $d_1+d_2>0$. Once $b<0$, the scalar field
condenses, that is, develops nonzero vacuum expectation value. For $d_2<0$ the
ground state has the form $\vek\phi_0=v(0,0,1)^T$, where
$v^2=-\frac{b}{2(d_1+d_2)}$. We will call this the \emph{polar phase} in analogy
with the three-color spin-one color superconductor. For $d_{2}>0$ the
ground state can be chosen as $\vek\phi_0=\frac{v}{\sqrt2}(1,\imag,0)^T$ with
$v^2=-\frac{b}{2d_1}$. This is analogous to the \emph{A-phase}.

In both phases, the $\gr{SO(3)}\times\gr{U(1)}$ global symmetry is broken to
a $\gr{U(1)'}$ subgroup. In the polar phase, this simply corresponds to
rotations in the $(\phi_1,\phi_2)$ plane, while in the A-phase, one has to
change the overall phase simultaneously with the rotation to keep the order
parameter unchanged. In the polar phase, all Noether charges of the global
$\gr{SO(3)}$ symmetry are zero in the ground state. As a consequence, we
expect to find three NG bosons with linear dispersion relations at low momentum,
associated with the three spontaneously broken generators. On the other hand, in
the A-state the third component of spin, represented in the $\vek\phi$-space by
the matrix $(J_i)_{jk}=-\imag\eps_{ijk}$, has nonzero density. Since $J_3$
belongs to a non-Abelian symmetry group, the NG boson counting will be modified
\cite{Nielsen:1975hm,Brauner:2005di}. We expect one type-I NG boson with a
linear dispersion relation and one type-II NG boson with a quadratic dispersion
relation. The latter represents a circularly polarized spin wave, very much
like in nonrelativistic ferromagnets. We will now see how these predictions,
based on general properties of spontaneous symmetry breaking in many-body
systems \cite{Brauner:2010wm}, are verified in an explicit calculation.


\subsection{Polar phase}
\label{Subsec:1color_polar}

The unbroken $\gr{U(1)'}\sim\gr{SO(2)}$ subgroup is generated by $J_3$ while the broken
generators are $J_{1,2}$ and the generator of phase transformations. The
six-(real-)component complex field $\vek\phi$ can therefore be parameterized as
\begin{equation}
\vek\phi=e^{\frac\imag v\theta}e^{\frac\imag v\vek\pi\cdot\vek
J}(\vek\phi_0+\vek H+\imag\vek\chi),
\end{equation}
where $\theta$ and $\vek\pi\equiv(\pi_1,\pi_2,0)^T$ are the NG modes, whereas
the remaining degrees of freedom, $\vek\chi=(-\chi_2,\chi_1,0)^T$ and
$\vek H=(0,0,H)^T$, are anticipated to excite massive states in the spectrum.
The parameterization of the vector $\vek\chi$ is just for convenience: to first
order in the fields the vector $\imag\vek\chi$ is then ``$\imag$ times
$\vek\pi$''. Plugging this parameterization into the Lagrangian
\eqref{Lagrangian_1color}, expanding up to second order in the fields, and
throwing away all total derivatives, we end up with the bilinear Lagrangian
\begin{eqnarray}
\La_{\text{bilin}}&=&-2c_1(H\de_0\theta-\vek\chi\cdot\de_0\vek\pi)+
\sum_\mu\binom{\mu=0:c_2}{\mu=i:-a_1}\left[(\de_\mu\theta)^2+(\de_\mu\vek\pi)^2+
(\de_\mu H)^2+(\de_\mu\vek\chi)^2\right]\nonumber \\
&&-a_2\left[(\de_3H-\text{rot}_\perp\vek\pi)^2+(\de_3\theta-\text{rot}_\perp\vek\chi)^2\right]
-4(d_1+d_2)v^2H^2+4d_2v^2\vek\chi^2,
\end{eqnarray}
where the planar rotation operator is defined as
$\text{rot}_\perp\vek\varphi\equiv\de_1\varphi_2-\de_2\varphi_1$.

Note that only the $H$ and $\vek\chi$ modes have mass terms as expected,
moreover,
the $\vek\chi$ mass vanishes for $d_2=0$. This signals the
instability associated with the phase transition from the polar to the A-phase.
On the other hand, the $H$ mass term can never change sign since $d_1+d_2>0$
is required by the stability of the potential. All six modes are in general
mixed by the derivative terms so that a direct diagonalization of the Lagrangian
would have to be done numerically. However, given that spatial rotations are
spontaneously broken to the $\gr{SO(2)}$ subgroup, we can investigate separately
excitations along the symmetry-breaking ($\phi_3$) axis, and in the
``transverse'' $(\phi_1,\phi_2)$ plane. The analysis then simplifies
considerably and essentially reduces to mixing of two fields at a time. This is
discussed in general in \ref{App:mixing}.

Let us first assume that the fields depend just on the time and the third
coordinate. The complicated mixing in the $a_2$ term then reduces to a mere
modification of the phase velocities of the $H$ and $\theta$ fields. Only the
singlets $H$ and $\theta$, and vectors $\vek\chi$ and $\vek\pi$ now mix.
The dispersion relations are found using Eqs.~\eqref{disp_Higgs} and
\eqref{disp_NGB}. In the $(H,\theta)$ sector they read
\begin{equation}
E^2=\frac1{c_2^2}[4v^2(d_1+d_2)c_2+c_1^2]\quad\text{(massive mode)},
\qquad
E^2=\frac{4v^2(d_1+d_2)(a_1+a_2)}{4v^2(d_1+d_2)c_2+c_1^2}k_3^2\quad
\text{(NG mode)},
\end{equation}
to the lowest order in the momentum expansion, while in the
$(\vek\chi,\vek\pi)$ sector they are
\begin{equation}
E^2=\frac1{c_2^2}(-4v^2d_2c_2+c_1^2)\quad\text{(massive modes)},
\qquad
E^2=\frac{-4v^2d_2a_1}{-4v^2d_2c_2+c_1^2}k_3^2\quad\text{(NG modes)}.
\label{polar_disp_long}
\end{equation}

Next we assume that the fields depend just on time and the first two
coordinates. The $H$ and $\theta$ modes then decouple from others, giving
rise to dispersion relations
\begin{equation}
E^2=\frac1{c_2^2}[4v^2(d_1+d_2)c_2+c_1^2]\quad\text{(massive mode)},
\qquad
E^2=\frac{4v^2(d_1+d_2)a_1}{4v^2(d_1+d_2)c_2+c_1^2}k_\perp^2\quad
\text{(NG mode)},
\end{equation}
where the subscript ``$\perp$'' indicates that the mode is transverse. On the
other hand, the $a_2$ term still seemingly mixes the components of the
$\vek\chi$ and $\vek\pi$ vectors. However, redefining them to
$\tilde{\vek\chi}=(\chi_1,\chi_2,0)^T$ and
$\tilde{\vek\pi}=(\pi_2,-\pi_1,0)^T$, one gets
$\text{rot}_\perp\vek\chi=\vek\de\cdot\tilde{\vek\chi}$ and
$\text{rot}_\perp\vek\pi=\vek\de\cdot\tilde{\vek\pi}$, while the $c_1$ term is not
affected since $\tilde{\vek\chi}\cdot\tilde{\vek\pi}=\vek\chi\cdot\vek\pi$. The
excitations in the $(\vek\chi,\vek\pi)$ sector then further split into two
branches. Modes for which $\tilde{\vek\chi}$ and $\tilde{\vek\pi}$ are parallel
to the momentum behave as longitudinal and their dispersion relations are given
by
\begin{equation}
E^2=\frac1{c_2^2}(-4v^2d_2c_2+c_1^2)\quad\text{(massive mode)},
\qquad
E^2=\frac{-4v^2d_2(a_1+a_2)}{-4v^2d_2c_2+c_1^2} k_\perp^2\quad
\text{(NG mode)}.
\end{equation}
For $\tilde{\vek\chi}$ and $\tilde{\vek\pi}$ modes perpendicular to the
momentum, the $a_2$ term vanishes and the dispersion relations are identical to
those found in Eq.~\eqref{polar_disp_long}.

\cor{}{Note that the masses of the massive modes naturally do not depend on the chosen direction of
momentum, as could have been expected. Also, it is now obvious that nonzero momentum lifts
degeneracy based on the symmetry of the ground state. Based on the unbroken $\gr{SO(2)}$ symmetry
we would have expected both $\vek\chi$ and $\vek\pi$ to transform as vectors whereas $H$ and
$\theta$ as singlets. This is indeed the case for the longitudinal excitations. On the other hand,
transverse momentum breaks the remaining $\gr{SO(2)}$ symmetry so that the $\pi_1,\pi_2$ NG modes
are no longer degenerate. A similar remark applies to all other results in this and the following
section.}


\subsection{A-phase}
\label{Subsec:1color_A}

The ground state is an eigenvector of $J_3$, that is,
$J_3\vek\phi_0=\vek\phi_0$.
This implies that the unbroken subgroup is generated
by the combination $\frac12(\openone-J_3)$. The broken generators can be
conveniently chosen as $J_1$, $J_2$, and $\frac12(\openone+J_3)$. So
$\vek\phi$ can be parameterized as
\begin{equation}
\vek\phi=e^{\frac\imag v\vek\pi\cdot\vek J'}\left(\vek\phi_0+\frac
Hv\vek\phi_0+\chi\vek\phi_1\right),
\end{equation}
where $\vek J'=\bigl(J_1,J_2,\frac{\openone+J_3}2\bigr)$,
$\vek\pi=(\pi_1,\pi_2,\pi_3)^T$,
and $\vek\phi _1=\frac1{\sqrt2}(1,-\imag,0)^T$ is a vector perpendicular to
$\vek\phi_0$. The real field $H$ and the complex field $\chi$ describe the
non-NG modes. When used in the Lagrangian \eqref{Lagrangian_1color}, this
parameterization yields the following bilinear terms
\begin{eqnarray}
\La_{\text{bilin}}&=&\imag c_1\left[
\chi^*\de_0\chi+2\imag H\de_0\pi_3+
\frac\imag2(\pi_1\de_0\pi_2-\pi_2\de_0\pi_1)\right]\nonumber\\
&&+\sum_\mu\binom{\mu=0:c_2}{\mu=i:-a_1}\left[
\frac12(\de_\mu\pi_{1,2})^2+(\de_\mu\pi_3)^2+
(\de_\mu H)^2+|\de_\mu\chi|^2\right] \nonumber\\
&&-\frac{a_2}2\left[(\de_3\pi_2-\de_2\pi_3+\de_1\chi_1+\de_2\chi_2+\de_1H)^2
+(\de_3\pi_1-\de_1\pi_3+\de_2\chi_1-\de_1\chi_2-\de_2H)^2\right]\nonumber\\
&&-4d_1v^2H^2-4d_2v^2|\chi|^2,
\end{eqnarray}
where $\chi_{1,2}$ are the real and imaginary parts of $\chi$. Again, thanks to
the exponential parameterization the NG fields are explicitly eliminated from
the static part of the Lagrangian. Also, the $\chi$ mass term is seen to vanish
at the transition to the polar phase, i.e., $d_2=0$.

To determine the excitation spectrum, we will first assume that the
fields depend only on time and the third coordinate. The fields then
fall into three sectors that do not mix with each other. The field
$\chi$ carries nonzero charge of the unbroken $\gr{U(1)'}$ symmetry
and describes a particle--antiparticle pair with masses
\begin{equation}
E=\frac 1{2c_2}\left(\pm c_1+ \sqrt{16v^2d_2c_2+c_1^2}\right).
\label{A_massive}
\end{equation}
The $H$ and $\pi_3$ modes give rise, similarly to the polar phase,
to one massive and one NG state of type-I,
\begin{equation}
E^2=\frac1{c_2^2}(4v^2d_1c_2+c_1^2)\quad\text{(massive mode)},
\qquad
E^2=\frac{4v^2d_1a_1}{4v^2d_1c_2+c_1^2}k_3^2\quad\text{(NG mode)}.
\end{equation}
On the contrary, the $(\pi_1,\pi_2)$ sector, which naively contains two NG
bosons, only produces one NG particle of type-II, in agreement with the general
discussion in \cite{Brauner:2005di},
\begin{equation}
E=\frac{|c_1|}{c_2}\quad\text{(massive mode)},
\qquad
E=\frac{a_1+a_2}{|c_1|}k_3^2\quad\text{(type-II NG mode)}.
\end{equation}

Second, we will investigate the transverse excitations. \cor{}{The masses of the particles excited
by $\chi_1,\chi_2$ are still given by Eq.~\eqref{A_massive}; the anisotropy brought by nonzero
momentum does not appear in this lowest-order term in the dispersion relation.}
For the $(H,\pi_3)$ fields we obtain the following dispersions,
\begin{equation}
E^2=\frac 1{c_2^2}(4v^2d_1c_2+c_1^2)\quad\text{(massive mode)},
\qquad
E^2=\frac{4v^2d_1\left(a_1+\frac{a_2}2\right)}{4v^2d_1c_2+c_1^2}
k_\perp^2 \quad\text{(type-II NG mode)}.
\end{equation}
The $\pi_1$ and $\pi_2$ modes trivially decouple and give rise to
the dispersions
\begin{equation}
E=\frac{|c_1|}{c_2}\quad\text{(massive mode)},
\qquad
E=\frac{a_1}{|c_1|} k_\perp^2\quad\text{(type-II NG mode)}.
\end{equation}

Before we conclude the section we remark that the calculation of the
dispersion relations is complicated by the fact that spacetime
symmetry is spontaneously broken. However, the basic anticipated
features of the NG spectrum are preserved. In both phases three
generators are spontaneously broken. In the polar phase, they give
rise to three type-I NG bosons with linear dispersion relation at
low momentum. In the A-phase, the ground state carries nonzero spin
density, therefore the three broken generators produce one type-I NG
boson with linear dispersion and one type-II NG boson with quadratic
dispersion at low momentum.

Finally, at the transition point between the two phases $d_2=0$, the
static part of the Lagrangian has an extended $\gr{SO(6)}$ symmetry
under which the polar and A-phase order parameters are degenerate.
Five of its generators are broken, leaving an $\gr{SO(5)}$ invariant
subgroup. This extended symmetry is also reflected in the NG
spectrum. For instance, in the polar phase the phase velocities of
two of the NG bosons go to zero so that their dispersions are
quadratic. This is in accordance with the general Nielsen--Chadha
counting rule \cite{Nielsen:1975hm}. However, note that this
extended symmetry is explicitly broken by the $c_1$ and $a_2$ terms
in the Lagrangian \eqref{Lagrangian_1color}. The extra NG bosons are
therefore only present in the classical theory, they will acquire
nonzero masses via radiative corrections \cite{Weinberg:1972fn}.

\section{Single-flavor spin-one color superconductor}
\label{Sec:1flavor} In this section we will study the spin-one color
superconductor that involves pairing of quarks of
three colors but a single flavor. The diquark condensate or the
order parameter $\bm{\Delta}$ is then a color
antitriplet and spin triplet, so it is a $3\times3$ complex matrix
and transforms as
\begin{eqnarray}
\bm{\Delta} & \rightarrow & U\bm{\Delta}R
\end{eqnarray}
where $U=\exp(i\theta_{a}\lambda_{a})\in\gr{U(3)_L}
=\gr{SU(3)_c}\times\gr{U(1)_B}$ and
$R=\exp(i\alpha_{i}J_{i})\in\gr{SO(3)_R}$ are transformation
matrices. Here $\lambda_{a}$ are eight Gell-Mann matrices and
$\lambda_{0}\equiv\sqrt{\frac{2}{3}}\openone$,
$(J_{i})_{jk}=-i\eps_{ijk}$ are generators of
$\gr{SO(3)_R}$, $\theta_{a}$ ($a=0,\dotsc,8$) and $\alpha_{i}$
($i=1,2,3$) are rotation angles in $\gr{U(3)_L}$ and
$\gr{SO(3)_R}$ group space. There are 18 real parameters in
$\bm\Delta$, among which 12 parameters are carried by the
$\gr{U(3)_L}\times\gr{SO(3)_R}$ transformation making a
12-dimensional degenerate vacuum manifold. Then
$\bm{\Delta}$ can be parameterized by the remaining 6 real parameters which
characterize different vacuum states \cite{Brauner:2008ma},
\begin{eqnarray}
\bm{\Delta} & = & \left(\begin{array}{ccc}
\Delta_{1} & i\delta _{3} & -i\delta _{2}\\
-i\delta _{3} & \Delta_{2} & i\delta _{1}\\
i\delta _{2} & -i\delta _{1} &
\Delta_{3}\end{array}\right).\label{eq:para}\end{eqnarray}

\subsection{Ginzburg--Landau free energy and ground states}
\label{Subsec:1flavor_PD}
The GL analysis is similar to that for superfluid Helium 3
\cite{Vollhardt:1990vw}. Up to fourth order in $\bm{\Delta}$ and two
derivatives, the most general $\gr{U(3)_L}\times\gr{SO(3)_R}$ \cor{}{and parity}
invariant Ginzburg--Landau free energy density functional can be written as
\begin{eqnarray}
\mathcal{F}[\bm{\Delta}]
& = & a_{1}\text{Tr}(\partial_{i}\bm{\Delta}\partial_{i}\bm{\Delta}^{\dagger})
+a_{2}(\partial_{i}\bm{\Delta}_{ai})(\partial_{j}\bm{\Delta}_{aj}^{*})
+b\text{Tr}(\bm{\Delta}\bm{\Delta}^{\dagger})
\nonumber \\
&&+d_{1}[\text{Tr}(\bm{\Delta}\bm{\Delta}^{\dagger})]^{2}
+d_{2}\text{Tr}(\bm{\Delta}\bm{\Delta}^{\dagger}\bm{\Delta}\bm{\Delta}^{\dagger}
)
+d_{3}\text{Tr}[\bm{\Delta}\bm{\Delta}^{T}(\bm{\Delta}\bm{\Delta}^{T})^{\dagger}].
\end{eqnarray}
\cor{}{The time-dependent GL functional, or Lagrangian, is then in general written as}
\begin{equation}
\mathscr{L}  =  ic_{1}
\text{Tr}[\bm{\Delta}^{\dagger}\partial_{0}\bm{\Delta}]+c_{2}
\textrm{Tr}[(\partial_{0}\bm{\Delta}^{\dagger})(\partial_{0}\bm{\Delta})] -
\mathcal{F}[\bm{\Delta}],
\end{equation}
\cor{}{see \ref{App:disp} for more details and the relation between the coefficients $c_1,c_2$.
The ground state is found} by minimizing $\mathcal{F}[\bm{\Delta}]$. The sign of $b$ determines
whether the order parameter is zero or nonzero. Hereafter we will assume that $b<0$. The phase
structure, or orientation in the field space, of $\bm{\Delta}_{0}$ depends on $d_{2}$ and $d_{3}$.
\cor{}{The magnitude of the condensate, $v=\sqrt{\text{Tr}(\bm\Delta\bm\Delta^\dagger)}$, }is given
by $v=\sqrt{-b/(2\bar{d})}$ with $\bar{d}=d_{1}+f(d_{2},d_{3})$ ($f$ is a function of $d_2$ and
$d_3$ that is specific to a particular phase) and the vacuum energy is
$E_{\text{vac}}=-b^{2}/(4\bar{d})$. The boundedness of the potential from below demands that $\bar
d>0$, which constrains the possible values of $d_1$ for a given phase.
The symmetry of the problem allows for altogether 8 inequivalent states
with different patterns of spontaneous breaking of the \emph{continuous}
symmetry. However, only the following four of them occupy a part of the phase
diagram \cite{Brauner:2008ma},
\begin{equation}
\label{eq:order}
\bm{\Delta}_{\text{CSL}}=\frac{1}{\sqrt{3}}\left(\begin{array}{ccc}
1 & 0 & 0\\
0 & 1 & 0\\
0 & 0 &
1\end{array}\right),\;\bm{\Delta}_{\text{polar}}=\left(\begin{array}{ccc}
0 & 0 & 0\\
0 & 0 & 0\\
0 & 0 &
1\end{array}\right),\;\bm{\Delta}_{\text{A}}=\frac{1}{\sqrt{2}}\left(
\begin{array}{ccc}
0 & 0 & 0\\
0 & 0 & 0\\
1 & i &
0\end{array}\right),\;\bm{\Delta}_{\varepsilon}=\left(\begin{array}{ccc}
0 & 0 & 0\\
0 & 0 & \beta\\
\alpha & i\alpha & 0\end{array}
\right),
\end{equation}
\cor{}{where $\alpha=\sqrt{(d_2+d_3)/[2(2d_2+d_3)]}$ and $\beta=\sqrt{d_2/(2d_2+d_3)}$.} 
The pattern of spontaneous symmetry breaking determines the low-energy spectrum of the system, i.e.,
the NG bosons. While some of the NG bosons are associated with the generators of the
color $\gr{SU(3)_c}$ group and are thus eventually absorbed in gluons via the Higgs--Anderson
mechanism, those stemming from spontaneous breaking of baryon number or rotation
symmetry remain in the spectrum as physical soft modes. As we will now see, some
of the phases exhibit the unusual type-II NG bosons, in accordance with general
properties of spontaneously broken symmetries in quantum many-body systems
\cite{Nielsen:1975hm,Brauner:2010wm}. \cor{}{We did not solve the fully coupled equations of motion
for the fields depending simultaneously on all coordinates, so the dispersion relations shown in the
following should be understood as combinations of separate formulas for the ``longitudinal'' and 
``transverse'' excitations.}

\subsection{CSL phase}
\label{Subsec:1flavor_CSL}
When $d_{2}+d_{3}>0$ and $d_{2}>d_{3}$,
the ground state is the CSL phase whose order parameter is given in
Eq. (\ref{eq:order}). Also, $\bar{d}=d_{1}+\frac{d_{2}+d_{3}}{3}$. In
the CSL phase the spin and color are coupled in the pairing, so the symmetry
breaking pattern is
$\gr{U(3)_L}\times\gr{SO(3)_R}\rightarrow\gr{SO(3)_V}$.
The generators of the unbroken symmetry $\gr{SO(3)_V}$ are
$\sqrt{\frac{1}{2}}(\lambda_{7}\otimes\openone+\openone\otimes J_{1})$,
$\sqrt{\frac{1}{2}}(-\lambda_{5}\otimes\openone+\openone\otimes J_{2})$,
$\sqrt{\frac{1}{2}}(\lambda_{2}\otimes\openone+\openone\otimes J_{3})$.
There are 9 broken generators leading to 9 NG bosons as follows,
\begin{itemize}
\item $\lambda_{0}\otimes \openone$. Type-I NG singlet, $E^{2}\sim
(a_{1}+a_{2})k^{2}$.
\item $\sqrt{\frac{1}{2}}(\lambda_{7}\otimes\openone-\openone\otimes J_{1})$
,$\sqrt{\frac{1}{2}}(-\lambda_{5}\otimes\openone-\openone\otimes J_{2})$,
$\sqrt{\frac{1}{2}}(\lambda_{2}\otimes\openone-\openone\otimes J_{3})$. Type-I NG
triplet, $E^{2}\sim(a_{1}+a_{2})k^{2}$.
\item \cor{}{$\lambda_{\alpha}\otimes \openone,\;\alpha=1,3,4,6,8$. Type-I NG 5-plet,
$E^{2}\sim(a_{1}+a_{2})k^{2}$.}
\end{itemize}
The details of the above modes are given in \ref{Subapp:disp_CSL}. After gauging
the color symmetry, the type-I NG 5-plet is absorbed by gluons, and we are left with the singlet as
the only physical NG boson, stemming from the spontaneous breaking of the $\gr{U(1)_B}$ symmetry.


\subsection{Polar phase}
\label{Subsec:1flavor_polar}
When $d_{3}<0$ and $d_{2}+d_{3}<0$,
the ground state is the polar phase whose order parameter is shown
in Eq. (\ref{eq:order}), and $\bar{d}=d_{1}+d_{2}+d_{3}$. The symmetry
breaking pattern is
$\gr{U(3)_L}\times\gr{SO(3)_R}\rightarrow\gr{U(2)_L}\times
\gr{SO(2)_R}$. The unbroken symmetry is generated by
$\lambda_{1,2,3}\otimes\openone$,
$\mathcal{P}_{12}\otimes\openone$ and $\openone\otimes J_{3}$,
where
$\mathcal{P}_{12}=\frac{1}{\sqrt{3}}(\sqrt{2}\lambda_{0}+\lambda_{8})=\mathrm{
diag}(1,1,0)$
is the projector onto the first two colors. The diquark spin is
polarized to one direction. \cor{}{There are 7 broken generators which, however, give rise only to
5 NG bosons, organized in the following multiplets,}
\begin{itemize}
\item $\sqrt{2}\mathcal{P}_{3}\otimes\openone$, where
$\mathcal{P}_{3}=\frac{1}{\sqrt{6}}(\lambda_{0}-\sqrt{2}\lambda_{8})=\mathrm{
diag}(0,0,1)$ is the projector onto the third color. Type-I NG singlet, $E^{2}\sim
a_{1}k_{\perp}^{2}+(a_{1}+a_{2})k_{3}^{2}$.
\item $\openone\otimes J_{j}$, $j=1,2$. Type-I NG doublet,
$E^{2}\sim(a_{1}+a_{2})k_{\perp}^{2}+a_{1}k_{3}^{2}$.
\item $\lambda_{\alpha}\otimes \openone$, $\alpha=4,5,6,7$. Type-II NG
doublet, $E^{2}\sim a_{1}^{2}k_{\perp}^{4}+(a_{1}+a_{2})^{2}k_{3}^{4}$.
\end{itemize}
\cor{}{The presence of type-II NG bosons is due to nonzero color density of the polar ground state.}
The details of the above modes are given in \ref{Subapp:disp_polar}. After
gauging the color symmetry, the type-I NG singlet and the type-II NG doublet are
absorbed by gluons, only the type-I NG doublet survives, corresponding to two
linearly polarized spin waves.

\subsection{A-phase}
\label{Subsec:1flavor_A}
When $d_{3}>0$ and $d_{2}<0$,
the ground state is the A-phase whose order parameter is shown in
Eq. (\ref{eq:order}), and $\bar{d}=d_{1}+d_{2}$. The symmetry breaking
pattern is
$\gr{U(3)_L}\times\gr{SO(3)_R}\rightarrow\gr{U(2)_L}\times
\gr{SO(2)_V}$.
The residual symmetry is generated by $\lambda_{1,2,3}\otimes\openone$,
$\mathcal{P}_{12}\otimes\openone$ and $\sqrt{\frac{2}{3}}
(\mathcal{P}_{3}\otimes\openone-\openone\otimes J_{3})$.
Unlike in the polar phase, the diquark spin is now circularly polarized.
Among 7 broken generators, there is only one giving rise to a type-I NG mode,
\begin{itemize}
\item \begin{flushleft}
$\sqrt{\frac{2}{3}}(\mathcal{P}_{3}\otimes\openone+\openone\otimes J_{3})$.
Type-I NG singlet, $E^{2}\sim(a_{1}+a_{2})k_{\perp}^{2}+a_{1}k_{3}^{2}$.
\par\end{flushleft}
\end{itemize}
The rest 6 generators produce only 3 type-II NG bosons due to non-zero
color and spin density of the A-phase vacuum,
\begin{itemize}
\item $\lambda_{\alpha}\otimes\openone$, $\alpha=4,5,6,7$. Type-II
NG doublet,
$E^{2}\sim(a_{1}+a_{2})^{2}k_{\perp}^{4}+a_{1}^{2}k_{3}^{4}$.
\item $\openone\otimes\frac{1}{\sqrt{2}}(J_{1}\pm iJ_{2})$. Type-II
NG singlet,
$E^{2}\sim a_{1}^{2}k_{\perp}^{4}+(a_{1}+a_{2})^{2}k_{3}^{4}$.
\end{itemize}
See \ref{Subapp:disp_A} for details on the above modes. After gauging the color
symmetry, the type-I NG singlet and the type-II NG doublet are absorbed by
gluons and leave the type-II NG singlet giving a circular spin wave.

\subsection{$\eps$-phase}
\label{Subsec:1flavor_eps}
When $d_{3}>d_{2}>0$,
the ground state is the $\varepsilon$ phase, see Eq. (\ref{eq:order})
for its order parameter; in this case
$\bar{d}=d_{1}+\frac{d_{2}(d_{2}+d_{3})}{2d_{2}+d_{3}}$. The symmetry breaking
pattern is
$\gr{U(3)_L}\times\gr{SO(3)_R}\rightarrow\gr{U(1)_L}\times
\gr{SO(2)_V}$. The spin of the second diquark color is longitudinally
polarized, while that of third color is circularly polarized. Like in
the A-phase, the circularly polarized spin produces an $\gr{SO(2)_V}$
unbroken symmetry. The unbroken symmetry is generated by
$\sqrt{2}\mathcal{P}_{1}\otimes\openone$
and $\sqrt{\frac{2}{3}}(\mathcal{P}_{3}\otimes\openone-\openone\otimes J_{3})$,
where $\mathcal{P}_{1}=\mathrm{diag}(1,0,0)$ is the projector onto
the first color. Out of the 10 broken generators only two correspond to type-I NG modes:
\begin{itemize}
\item $\sqrt{2}\mathcal{P}_{2}\otimes\openone$, where
$\mathcal{P}_{2}=\mathrm{diag}(0,1,0)$
is the projector onto the second color. Type-I NG singlet, $E^{2}\sim
a_{1}k_{\perp}^{2}+(a_{1}+a_{2})k_{3}^{2}$.
\item $\sqrt{\frac{2}{3}}(\mathcal{P}_{3}\otimes\openone+\openone\otimes J_{3})$.
Type-I NG singlet, $E^{2}\sim(a_{1}+a_{2})k_{\perp}^{2}+a_{1}k_{3}^{2}$.
\end{itemize}
The remaining 8 generators give rise to 4 type-II NG modes due to non-zero
color and spin density of the $\varepsilon$ vacuum,
\begin{itemize}
\item $\frac{1}{\sqrt{2}}(\lambda_{1}\pm i\lambda_{2})\otimes\openone$.
Type-II NG singlet,
$E^{2}\sim a_{1}^{2}k_{\perp}^{4}+(a_{1}+a_{2})^{2}k_{3}^{4}$.
\item $\frac{1}{\sqrt{2}}(\lambda_{4,6}\pm i\lambda_{5,7})\otimes\openone$.
Type-II NG doublet,
$E^{2}\sim (a_{1}+a_{2})^{2}k_{\perp}^{4}+a_{1}^{2}k_{3}^{4}$.
\item $\openone\otimes\frac{1}{\sqrt{2}}(J_{1}\pm iJ_{2})$. Type-II
NG singlet, $E^{2}\sim a_{1}^{2}k_{\perp}^{4}+(a_{1}+a_{2})^{2}k_{3}^{4}$.
\end{itemize}
See \ref{Subapp:disp_eps} for details on the above modes.
After gauging the color symmetry, only the last type-II NG singlet
survives, corresponding to a circularly polarized spin wave.

\section{Low energy effective field theory for the CSL phase}
\label{Sec:CSL_EFT} The CSL phase plays in many respects a
distinguished role among all spin-one phases investigated in the
preceding section. First, it is the ground state of one-flavor quark
matter in the limit of very high chemical potential. Second, it is
isotropic and involves democratically all quark colors.
Consequently, under some additional assumptions on the spin-orbital
structure of the order parameter, all quarks are gapped and the
low-energy spectrum is solely determined by the NG bosons of the
spontaneously broken symmetry. Moreover, the rotational symmetry is
unbroken whereas the NG bosons associated with the color symmetry
are absorbed into gluons once this is gauged. Therefore the
low-energy physics of the CSL phase will be governed by the NG
bosons of the global \emph{Abelian} symmetry. In this section, we
will employ a technique of Son \cite{Son:2002zn}, which was
previously used to study the transport properties of the CFL phase
\cite{Manuel:2004iv}.


\subsection{Global symmetry and NG bosons}
\label{Subsec:CSL_EFT_symmetry} Based on the argument in the
previous paragraph, one would naively conclude that in the
one-flavor CSL phase, there is exactly one physical NG boson,
stemming from the spontaneous breaking of the global $\gr{U(1)}$
symmetry. However, this is only true provided the quarks do not
carry any other gauge degrees of freedom apart from the color ones.
This is certainly a very rough approximation considering that the
spin-one phases are only likely to occur in the phase diagram in the
region where strange quark mass is large and electric charge
neutrality effects play an important role.

As already discussed in the Introduction, there are two realistic
scenarios for spin-one pairing to occur in three-flavor quark matter
\cite{Alford:2007xm}. (i) The $u$ and $d$ quarks are paired in the
2SC phase and only the $s$ quarks are left and undergo the
single-flavor spin-one pairing. (ii) Cross-flavor pairing is
completely prohibited by large Fermi level mismatch and all three
quark flavors undergo single-flavor pairing. While in the case (ii)
the most favored pairing pattern will be CSL, in the case (i) it may
not be so. The reason is that the 2SC phase is not color neutral; to
compensate for its color charge, a color chemical potential must be
introduced. This in turn breaks the exact color symmetry of the $s$
quark sector. In addition, the mismatch between different colors may
favor other pairing patterns such as the polar one
\cite{Alford:2005yy}.

For these reasons, we will only consider the scenario (ii) which is
theoretically clean; each flavor feels an exact color $\gr{SU(3)}$ symmetry and
features the same symmetry breaking pattern. The question then is: how many NG
bosons are there? As mentioned above, the NG bosons of the spontaneously broken
color symmetry will be absorbed into gluons. In addition, there is a global
Abelian symmetry, $\gr G=\gr{U(1)}_u\times\gr{U(1)}_d\times\gr{U(1)}_s$,
corresponding simply to separate conservation of the flavor quark numbers. This
will give rise to three NG bosons. However, a subgroup of $\gr G$, given by the
electromagnetic $\gr{U(1)_Q}$, is gauged and the associated NG boson will be
eaten by the photon, making it massive and thus giving rise to the Meissner
effect. Therefore, there will be only two physical NG bosons left.

One should note that in special limits, the flavor symmetry can be
actually larger than $\gr G$. For example, assuming that the quark
masses satisfy $0\neq m_u=m_d\neq m_s$, the flavor symmetry will be
$\tilde{\gr G}=\gr{SU(2)_V}\times\gr{U(1)}_{u+d}\times\gr{U(1)}_s$.
However, the non-Abelian $\gr{SU(2)_V}$ group will be explicitly
broken down to $\gr{U(1)_{I_3}}$, generated by the third component
of isospin, by the electric charge chemical potential necessary to
maintain electric charge neutrality. The remaining exact symmetry
group is isomorphic to $\gr G$. The very fact that $u$ and $d$
quarks are actually light does not play a role since the pions will
presumably still be heavier than the CSL gap so that they will not
enter the low-energy effective field theory whose validity is
limited by the scale of the gap.


\subsection{General method to construct the effective Lagrangian}
\label{Subsec:CSL_EFT_general} A general method to construct the
low-energy effective action for the NG boson of a spontaneously
broken $\gr{U(1)}$ symmetry at zero temperature was proposed by Son
based on a functional technique \cite{Son:2002zn}. Starting with the equation
of state, $P(\mu)$, the effective action for the NG field $\vp$ to
the lowest order in derivatives reads
\begin{equation}
\Gamma[\mu,\vp]=\int\mathrm{d}^4\!x\,P\Bigl(\sqrt{(\de_0\vp-\mu)^2-(\de_i\vp)^2}\Bigr).
\label{Son_EFT}
\end{equation}
Several remarks are in order here. First, this is a fully \emph{quantum}
effective action, that is, it should be used strictly at tree level. All loop
effects are included in the couplings of the action. Second, after
expansion in powers of $\vp$, the action contains only terms with the same
number of derivatives as is the power of $\vp$. This is the leading-order term
in the derivative expansion for a given power of $\vp$. Third, the derivation
relies heavily on the fact that $\mu$ is the only source of Lorentz violation in
the theory, since then the full dependence on the NG field can be reconstructed
from the dependence of the pressure on the chemical potential using the fact
that the NG field and the chemical potential only appear in the effective
action in the combination $\de_\nu\vp-\delta_{\nu0}\mu$.

Once we know the equation of state, we plug it into
Eq.~\eqref{Son_EFT} and expand in powers of $\vp$ to obtain both the
dispersion relation of the NG boson and its self-interactions. In
extremely dense quark matter one can, thanks to asymptotic freedom,
take as a reasonable starting point the equation of state for a free
massless Fermi gas $P_0(\mu)=N_c\mu^4/(12\pi^2)$, for a single quark
flavor \cite{Son:2002zn,Manuel:2004iv}. The effect of pairing on the
equation of state can be neglected since the CSL gap is numerically
very small. However, one may be interested in corrections due to
nonzero quark mass since it is exactly the strange quark mass that
opens the way to the spin-one phases in the phase diagram. To that
end, one needs to know the equation of state of a massive Fermi gas,
\begin{equation}
\frac{P_m(\mu)}{N_c}=\frac{\mu k_{\rm F}^3}{12\pi^2}-\frac{m^2\mu k_{\rm F}}{8\pi^2}+
\frac{m^4}{8\pi^2}\log\frac{\mu+k_{\rm F}}{m}
\approx\frac{\mu^4}{12\pi^2}-\frac{m^2\mu^2}{4\pi^2}+\mathcal O(m^4\log m),
\end{equation}
where $k_{\rm F}=\sqrt{\mu^2-m^2}$ is the Fermi momentum. Substituting this equation
of state in the general formula \eqref{Son_EFT}, one obtains the effective
Lagrangian
\begin{equation}
\begin{split}
\frac1{N_c}\La_{\text{eff}}(\vp)=&\frac{1}{12\pi^2}\left[\mu^4-4\mu^3\de_0\vp
+6\mu^2(\de_0\vp)^2-2\mu^2(\de_i\vp)^2-4\mu\de_0\vp\de_\mu\vp\de^\mu\vp
+(\de_\mu\vp\de^\mu\vp)^2\right]-\\
&-\frac{m^2}{4\pi^2}\left[\mu^2-2\mu\de_0\vp+(\de_0\vp)^2
-(\de_i\vp)^2\right].
\end{split}
\label{leff}
\end{equation}
The bilinear terms in the Lagrangian imply that the NG boson phase velocity
equals
\begin{equation}
v^2=\frac13\frac{2\mu^2-3m^2}{2\mu^2-m^2}\approx\frac13
\left(1-\frac{m^2}{\mu^2}\right)+\mathcal O(m^4/\mu^4).
\label{velocity}
\end{equation}
This can be shown to coincide, to the order displayed, with the
hydrodynamic speed of sound in a free gas.


\subsection{Effective Lagrangian for neutral quark matter}
\label{CSL_EFT_EFT}

Let us now consider the case of several $\gr{U(1)}$ symmetries with
different chemical potentials. Thus, the spontaneously broken
symmetry group $\gr{G=U(1)_1\times U(1)_2\times\dotsb}$ is
associated with the NG fields $\vp_1,\vp_2,\dotsc$, chemical
potentials $\mu_1,\mu_2,\dotsc$, and the equation of state
$P(\mu_1,\mu_2,\dotsc)$. Unfortunately, one can easily see that
Son's trick does not work in this case. The reason is that with more
fields there are more independent ways to construct a
Lorentz-invariant Lagrangian density that reduces to the same
function of chemical potentials for uniform fields. As an example,
just observe that $D^\mu\vp_1D_\mu\vp_1D^\nu\vp_2D_\nu\vp_2$ and
$D^\mu\vp_1D^\nu\vp_1D_\mu\vp_2D_\nu\vp_2$ both give
$\mu_1^2\mu_2^2$.

Fortunately, there is a special case where Son's method can still be used.
Once the equation of state separates to
\begin{equation}
P(\mu_1,\mu_2,\dotsc)=P_1(\mu_1)+P_2(\mu_2)+\dotsb,
\label{eos}
\end{equation}
the dependence of the effective action on the NG fields $\vp_k$ can
again be fully reconstructed using Lorentz invariance. This
generalization of the method may seem somewhat trivial, since the
equation of state \eqref{eos} corresponds to separate and
noninteracting subsystems. However, they can become entangled by a
coupling to an additional field, as we will see later.

Let us address the specific question: what happens when a part of
the symmetry group is gauged? For simplicity we will assume that
there is only one gauge field that couples to a linear combination
of generators of $\gr{U(1)}_k$, that is, to a subgroup of $\gr G$.
This is determined by the charges $q_k$ of the fields $\vp_k$. In
the effective Lagrangian we thus have to replace the combinations
$\de_\nu\vp_k-\delta_{\nu0}\mu_k$ with
$D_\nu\vp_k-\delta_{\nu0}\mu_k$, where
$D_\mu\vp_k=\de_\mu\vp_k-eq_kA_\mu$ and $e$ is the gauge coupling.
To obtain the effective Lagrangian from the equation of state, one
in turn has to replace everywhere $\mu_k^2$ with
$(D_0\vp_k-\mu_k)^2-(D_i\vp_k)^2$. Note that the effective
Lagrangian also contains a term $A_0J_0$: a coupling of the gauge
field to an external background charge density which ensures that
the system as a whole remains neutral despite the chemical
potentials $\mu_k$ \cite{Kapusta:1981aa,Gusynin:2003yu}. This is
equivalently expressed by the fact that $\langle A_0\rangle=0$.

We will now consider a system where the underlying equation of state
is well approximated by a noninteracting Fermi gas. This is the case
of color superconductors at high baryon density since the pairing
effects are exponentially suppressed and the normal Fermi liquid
contribution dominates the pressure. In accordance with
Eq.~\eqref{leff}, the effective Lagrangian with the leading
finite-mass correction then reads
$\La_{\text{eff}}=N_c(\La_0+\La_1)+\La_{\rm g}$, where (upon
omitting terms of zeroth and first order in the fields)
\begin{equation}
\begin{split}
\La_0=&\frac1{12\pi^2}\sum_k\left[
6\mu_k^2(D_0\vp_k)^2-2\mu_k^2(D_i\vp_k)^2
-4\mu_kD_0\vp_kD_\mu\vp_kD^\mu\vp_k+(D_\mu\vp_kD^\mu\vp_k)^2\right],\\
\La_1=&-\sum_k\frac{m_k^2}{4\pi^2}\left[(D_0\vp_k)^2-(D_i\vp_k)^2\right],\qquad
\La_{\rm g}=-\frac14F_{\mu\nu}F^{\mu\nu}.
\end{split}
\label{L0}
\end{equation}

So far we have not fixed the gauge for $A_\mu$.
Note that the covariant derivative $\partial_{\mu}\varphi_{k}-eq_{k}A_{\mu}$
is invariant under the gauge transformation, $\varphi'_{k} = \varphi_{k}+eq_{k}\alpha$,
$A'_{\mu} = A_{\mu}+\partial_{\mu}\alpha$, where $\alpha$ is a gauge parameter.
We can change variables $\varphi _k$ to $\theta _k$
by $\varphi_{k}=R_{k\ell}\theta_{\ell}$, where $R_{k\ell}$ is a real
square matrix such that $R_{k1}\sim q_k$ up to a common factor (the detailed
form of this matrix will be specified later). The gauge transformation reads
$R\theta'=R\theta+e\alpha q$ or $\theta'=\theta+e\alpha R^{-1}q$ where
$R,R^{-1}$ are matrices and $\theta',\theta,q$ vectors. We can fix the gauge by choosing the gauge parameter
$\alpha=-\frac{\theta_{1}}{e(R^{-1})_{1k}q_{k}}$ so that $\theta'_{1}=0$.
We see that by fixing the gauge we can remove the field $\theta_1$, or we can set $\theta_1=0$.
All other modes $\theta_2,\theta_3,\dotsc$ remain
in the theory as physical fields. The resulting Lagrangian will be rather
complicated, let us therefore look explicitly at least at the bilinear part of
$\La_0+\La_1$,
\begin{equation}
\begin{split}
\La_{\text{bilin}}=\sum_k\left[\frac{2\mu_k^2-m_k^2}{4\pi^2}(D_0\vp_k)^2
-\frac{2\mu_k^2-3m_k^2}{12\pi^2}(D_i\vp_k)^2\right].
\end{split}
\label{eftgeneral}
\end{equation}
Expanding the square of the covariant derivative one obtains (no summation over
$k$ or $\mu$ is implied)
\begin{equation}
(D_\mu\vp_k)^2=(R_{k\ell}\de_\mu\theta_\ell-eq_kA_\mu)^2=
R_{k\ell}R_{km}\de_\mu\theta_\ell\de_\mu\theta_m-
2eq_kR_{k\ell}A_\mu\de_\mu\theta_\ell+e^2q_k^2A_\mu A_\mu.
\end{equation}

We can see that the gauge boson acquires a mass term, as expected.
Whether this affects the low-energy dynamics of the system is a
matter of scales. In order to have a clear physical interpretation
of the excitation spectrum, it would be better to get rid of the
mixing term $A_\mu\de_\mu\theta_\ell$. One could find the dispersion
relations even with such mixing, but most likely just numerically
\cite{Gusynin:2003yu}. Another,
practical aspect is that once we decide to integrate out the massive gauge boson
to obtain an effective Lagrangian for the NG bosons only, in the absence of the
mixing this can be done perturbatively and will merely result in a modification
of the NG boson interactions. On the other hand, the mixing with the gauge boson
would necessarily induce corrections to the NG boson dispersion relations.
Unfortunately it seems that in the most general case of Eq.~\eqref{eftgeneral},
the mixing term cannot be removed by a judicious choice of $R_{k\ell}$
simultaneously in the temporal and spatial parts of $\La_{\text{bilin}}$.
However, there are special cases in which the Lagrangian can be further
simplified.

(i) \emph{Zero masses.}
In this case, the NG--gauge boson mixing can be removed by choosing $R_{k\ell}$
so that $\sum_k\mu_k^2q_kR_{k\ell}=0$ for all $\ell=2,3,\dotsc$. This means that
the second and other columns of $R_{k\ell}$ should be set orthogonal to the
vector $\mu_k^2q_k$, that is, not to the linear combination that defines
$\theta_1$. It is more convenient to define $\tilde R_{k\ell}=\mu_kR_{k\ell}$
since the above condition then demands that the $\ell=2,3,\dotsc$ columns of
$\tilde R_{k\ell}$ be orthogonal to the first one. We can then choose the whole
matrix $\tilde R_{k\ell}$ to be orthogonal, upon which the bilinear Lagrangian
\eqref{eftgeneral} becomes
\begin{equation}
\La_{\text{bilin}}=\frac{1}{2\pi^2}\biggl\{
\sum_{\ell\neq1}\Bigl[(\de_0\theta_\ell)^2-\frac13(\de_i\theta_\ell)^2\Bigr]+
\sum_ke^2q_k^2\mu_k^2\Bigl(A_0A_0-\frac13A_iA_i\Bigr)\biggr\}.
\end{equation}
The interaction terms are obtained upon expressing $\vp_k$ in terms of
$\theta_\ell$ in Eq.~\eqref{L0}.

(ii) \emph{Equal masses and chemical potentials.} In this (rather unphysical)
case the mixing term is removed by setting $\sum_kq_kR_{k\ell}=0$ for all
$\ell=2,3,\dotsc$. We can thus choose directly the matrix $R_{k\ell}$ as
orthogonal and the resulting bilinear Lagrangian reads
\begin{equation}
\La_{\text{bilin}}=\frac{2\mu^2-m^2}{4\pi^2}\biggl\{
\sum_{\ell\neq1}\left[(\de_0\theta_\ell)^2-v^2(\de_i\theta_\ell)^2\right]
+e^2\sum_kq_k^2\left(A_0A_0-v^2A_iA_i\right)\biggr\},
\end{equation}
where the phase velocity $v$ is given by Eq.~\eqref{velocity}.

All the general formulas above are easily applied to the case of interest, that
is, three-flavor quark matter with the electric charge neutrality constraint.
Then the gauged subgroup is $\gr{U(1)}_Q$, generated by the electric
charge operator, $Q=(2/3,-1/3,-1/3)$. While the $u$ and $d$ quark masses can
certainly be neglected, the strange quark mass must be taken into account,
at least for the very reason that without this mass, no mismatch between the
Fermi momenta of different quark flavors would arise and the ground state would
be the CFL phase.


\section{Conclusions}
\label{Sec:conclusions}

We analyzed the low-energy physics of spin-one color superconductors in
terms of their soft, NG excitations. We used the Ginzburg--Landau theory to
derive the excitation spectrum. As a warm-up exercise we first analyzed pairing
of quarks of two flavors and a single color, which may play a role in
phenomenology as a complement to the 2SC pairing. Already this simple example
exhibits a phase with an unusual number of NG bosons. In particular, one of the
NG bosons in the A-phase is of type-II, i.e., has a quadratic dispersion
relation at low momentum, very much like the spin waves in ferromagnets.
Also, thanks to the fact that a spacetime (rotational) symmetry is
spontaneously broken we observed that the classification of NG modes into
multiplets of unbroken symmetry holds strictly only in the long-wavelength
limit. Any nonzero momentum of the soft mode further breaks the symmetry and
lifts the degeneracy based on the symmetry of the ground state itself.

The main body of the paper is comprised of an investigation of a
single-flavor three-color superconductor. This is the most likely candidate
phase for the ground state of dense quark matter in case that strange-quark
mass and electric charge neutrality effects disfavor pairing of quarks of
different flavors. We have thus completed the analysis of the phase diagram
started in Ref.~\cite{Brauner:2008ma} and the classification of the soft modes,
sketched in our previous paper \cite{Brauner:2009df}. The four phases that
appear in the phase diagram possess a plethora of different NG modes of the
spontaneously broken color, baryon number and rotational symmetry. Those
stemming from the color symmetry will eventually be absorbed into gluons,
making them massive by the Anderson--Higgs mechanism. The other NG bosons will
remain in the spectrum as physical soft modes.

Unlike in all the other phases, in the isotropic CSL phase all quarks
can be gapped so that the NG bosons are the only truly gapless states in the
spectrum. This has far-reaching consequences for the low-energy dynamics of the
CSL phase such as its transport properties. We laid the foundation for a later
economical calculation of the transport coefficients of the CSL phase such as
the shear viscosity by deriving the low-energy effective field theory for its
physical NG bosons. We used a functional technique \cite{Son:2002zn} applied to
the CFL phase before \cite{Manuel:2004iv} and adapted it for the present case by
introducing several independent chemical potentials and gauging a subgroup of
the symmetry group, corresponding to the electric charge. The actual
calculation goes beyond the scope of this paper and will be a subject of our
future work.


\section*{Acknowledgments}
The authors are grateful to X.-g.~Huang for useful discussions on the material of
Sec.~\ref{Sec:CSL_EFT}. The work of T.B. was supported in part by the ExtreMe Matter Institute EMMI
in the framework of the Helmholtz Alliance Program of the Helmholtz Association (HA216/EMMI), and
by the Sofja Kovalevskaja program of the Alexander von Humboldt Foundation. Q.W. is supported in
part by the `100 talents' project of Chinese Academy of Sciences (CAS) and by the National Natural
Science Foundation of China (NSFC) under the grant
10735040.


\appendix
\section{Mixing Lagrangians}
\label{App:mixing}

Similar to what has been done in \cite{Brauner:2005di}, let us consider the
mixing Lagrangian for two fields $\pi,H$ of the form
\begin{equation}
\La_{\text{mixing}}=\frac12\left[(\de_0\pi)^2-v^2(\vek\de\pi)^2\right]+
\frac12\left[(\de_0H)^2-v^2(\vek\de H)^2\right]-\frac12m^2H^2-\xi H\de_0\pi.
\label{mixing_Lagrangian}
\end{equation}
Here $v$ is the phase velocity; it will be sufficient to assume that it is
common to both fields, even though the result is easy to generalize to the case
with two different phase velocities. The field $H$ possibly has a mass term and
$\xi$ is the mixing parameter which arises from the term in Eq.
\eqref{Lagrangian_1color} with a single time derivative. The dispersion
relations following from the Lagrangian \eqref{mixing_Lagrangian} are
\begin{align}
\label{disp_Higgs}
E^2&=m^2+\xi^2+\mathcal O(\vek k^2),&&\text{massive mode,}\\
\label{disp_NGB}
E^2&=\frac{m^2v^2}{m^2+\xi^2}\vek k^2+\frac{\xi^4v^4}{(m^2+\xi^2)^3}\vek
k^4+\mathcal O(\vek k^6),&&\text{NG mode.}
\end{align}
Note in particular that when the mass term $m$ is zero, the mixing term $\xi$
transforms the two expected (type-I) NG modes into one massive mode with mass
$|\xi|$ and one (type-II) NG mode with the quadratic dispersion relation
$E=v^2\vek k^2/|\xi|$.


\section{Lagrangian and dispersion relations for spin-one color superconductor}
\label{App:disp}
The low-energy effective Lagrangian for the superconductor is in general only constrained by
rotational invariance as well as the internal symmetry of the system and, in case of QCD,
conservation of parity. Its most general form for the single-flavor spin-one color superconductor
therefore reads 
\begin{eqnarray}
\mathscr{L} & = & \imag c_{1} \text{Tr}[\bm{\Delta}^{\dagger}\partial_{0}\bm{\Delta}]+c_{2}
\textrm{Tr}[(\partial_{0}\bm{\Delta}^{\dagger})(\partial_{0}\bm{\Delta})]
 -a_{1}\text{Tr}(\partial_{i}\bm{\Delta}\partial_{i}\bm{\Delta}^{\dagger})
-a_{2}(\partial_{i}\bm{\Delta}_{ai})(\partial_{j}\bm{\Delta}_{aj}^{*})\nonumber\\
&& -b\text{Tr}(\bm{\Delta}\bm{\Delta}^{\dagger})
-d_{1}[\text{Tr}(\bm{\Delta}\bm{\Delta}^{\dagger})]^{2}
-d_{2}\text{Tr}(\bm{\Delta}\bm{\Delta}^{\dagger}\bm{\Delta}\bm{\Delta}^{\dagger})
-d_{3}\text{Tr}[\bm{\Delta}\bm{\Delta}^{T}(\bm{\Delta}\bm{\Delta}^{T})^{\dagger}].
\label{lag-csc-1}
\end{eqnarray}
In the limit that the Cooper pairs form tightly bound molecules, the ground state behaves as their
Bose--Einstein condensate. The coefficients $c_1$ and $c_2$ are then related. To see this, note
that adding a kinetic term to the free energy functional $\mathcal{F}[\bm{\Delta}]$ can be
understood as defining a Hamiltonian that governs the dynamics of the molecules, $\mathscr
H=c_{2}\textrm{Tr}[(\partial_{0}\bm{\Delta}^{\dagger})(\partial_{0}\bm{\Delta})]+\mathcal{F}[\bm{
\Delta}]$. The many-body description of the system is then accomplished by adding the chemical
potential $\mu$ via $\mathscr{H}\rightarrow\mathscr{H}-\mu\mathscr{N}$ where $\mathscr{N}$ is the
Noether charge corresponding to the $\gr{U(1)_B}$ symmetry (baryon number operator). Integrating out
the canonical momenta in order to arrive at a Lagrangian formulation of the many-body problem
\cite{Kapusta:1981aa}, one finds terms with one as well as two time derivatives whose coefficients
are related by $c_1=-2\mu c_2$. Nevertheless, we will keep them as independent parameters.


\subsection{CSL phase}
\label{Subapp:disp_CSL} With the knowledge of the broken and unbroken symmetry generators, we can
write the order parameter
$\bm{\Delta}$ in the following form,
\begin{eqnarray}
\bm{\Delta} & = & \exp(i\theta)\exp\left(\frac{1}{2}i\kappa_{a}\lambda_{a}\right)
(v\bm{\Delta}_{\mathrm{CSL}}+H)\exp(i\nu_{i}J_{i}),\end{eqnarray}
where the summations over $a$ and $i$ are in the ranges $a=1,3,4,6,8$
and $i=1,2,3$. The matrix field $H$ can be parameterized by
\begin{eqnarray}
H & = & \cor{}{h\openone+\frac12\varphi_a\lambda_a+\chi_iJ_i} =  \left(\begin{array}{ccc}
h+\frac{1}{2}\varphi_{3}+\frac{1}{2\sqrt{3}}\varphi_{8} & \frac{1}{2}\varphi_{1}-i\chi_{3} & \frac{1}{2}\varphi_{4}+i\chi_{2}\\
\frac{1}{2}\varphi_{1}+i\chi_{3} & h-\frac{1}{2}\varphi_{3}+\frac{1}{2\sqrt{3}}\varphi_{8} & \frac{1}{2}\varphi_{6}-i\chi_{1}\\
\frac{1}{2}\varphi_{4}-i\chi_{2} & \frac{1}{2}\varphi_{6}+i\chi_{1} & h-\frac{1}{\sqrt{3}}\varphi_{8}\end{array}\right),\end{eqnarray}
where $h,\varphi_{1,3,4,6,8},\chi_{1,2,3}$ are all massive fields. We find the following dispersion
relations,
\begin{eqnarray}
E_{h}^{2} & = &
\frac{-2bc_{2}+c_{1}^{2}}{c_{2}^{2}}+\frac{(a_{1}+a_{2})(-2bc_{2}+2c_{1}^{2})}
{-2bc_{2}^{2}+c_{1}^{2}c_{2}}k^{2},\nonumber\\
E_{\chi}^{2} & = &
\frac{-2bc_{2}(d_{2}-d_{3})+3\bar{d}c_{1}^{2}}{3c_{2}^{2}\bar{d}}+\frac{2(a_{1}+a_
{2})[-bc_{2}(d_{2}-d_{3})+3\bar{d}c_{1}^{2}]}{-2bc_{2}^{2}(d_{2}-d_{3})+3\bar{d}c_{1}^{2}c_{2}}
k^{2},\nonumber\\
E_{\varphi}^{2} & = &
\frac{-2bc_{2}(d_{2}+d_{3})+3\bar{d}c_{1}^{2}}{3c_{2}^{2}\bar{d}}+\frac{2(a_{1}+a_
{2})[-bc_{2}(d_{2}+d_{3})+3\bar{d}c_{1}^{2}]}{-2bc_{2}^{2}(d_{2}+d_{3})+3\bar{d}c_{1}^{2}c_{2}}
k^{2},\nonumber\\
E_{\theta}^{2} & = &
\frac{-2b(a_{1}+a_{2})}{-2bc_{2}+c_{1}^{2}}k^{2}+\frac{(a_{1}+a_{2})^{2}c_{1}^{4}}{
(-2bc_{2}+c_{1}^{2})^{3}}k^{4},\nonumber\\
E_{\nu}^{2} & = &
\frac{-2b(a_{1}+a_{2})(d_{2}-d_{3})}{-2bc_{2}(d_{2}-d_{3})+3\bar{d}c_{1}^{2}}k^{2}
+\frac{27(a_{1}+a_{2})^{2}c_{1}^{4}\bar{d}^{3}}{[-2bc_{2}(d_{2}-d_{3})+3\bar{d}c_{1}^{2}]^{3}}
k^{4},\nonumber\\
E_{\kappa}^{2} & = &
\frac{-2b(a_{1}+a_{2})(d_{2}+d_{3})}{-2bc_{2}(d_{2}+d_{3})+3\bar{d}c_{1}^{2}}k^{2}
+\frac{27(a_{1}+a_{2})^{2}c_{1}^{4}\bar{d}^{3}}{[-2bc_{2}(d_{2}+d_{3})+3\bar{d}c_{1}^{2}]^{3}}
k^{4}.
\end{eqnarray}

\subsection{Polar phase}
\label{Subapp:disp_polar} The order parameter can be written in the
following form \begin{eqnarray} \bm{\Delta} & = &
\exp(i\theta)\exp\left(\frac{1}{2}i\kappa_{a}\lambda_{a}\right)
(v\bm{\Delta}_{\mathrm{polar}}+H)\exp(i\nu_{i}J_{i}),\end{eqnarray}
where the summations over $a$ and $i$ are in the ranges $a=4,5,6,7$
and $i=1,2$. The matrix field $H$ can be parameterized by
\begin{eqnarray}
H & = & \left(\begin{array}{ccc}
\chi_{11}+i\varphi_{11} & \chi_{12}+i\varphi_{12} & 0\\
\chi_{21}+i\varphi_{21} & \chi_{22}+i\varphi_{22} & 0\\
i\rho_{1} & i\rho_{2} & h\end{array}\right).
\end{eqnarray}
\cor{}{It comprises of two complex doublets of the unbroken $\gr{SU(2)_L}$, a real vector $\rho$ of
$\gr{SO(2)_R}$, and a singlet $h$. They are all expected to excite massive modes.}
The NG mode $\theta$ and the massive mode $h$ are coupled, their eigenmodes are found to be
\begin{eqnarray}
E_{h}^{2} & = & \left(\frac{-2b}{c_{2}}+\frac{c_{1}^{2}}{2c_{2}^{2}}\right)
+\left(1-\frac{c_{1}^{2}}{4bc_{2}-c_{1}^{2}}\right)\left(\frac{a_{1}}{c_{2}}k_{\perp}^{2}
+\frac{a_{1}+a_{2}}{c_{2}}k_{3}^{2}\right),\nonumber \\
E_{\theta}^{2} & =&
\frac{4bc_{2}}{4bc_{2}-c_{1}^{2}}\left(\frac{a_{1}}{c_{2}}k_{\perp}^{2}
+\frac{a_{1}+a_{2}}{c_{2}}k_{3}^{2}\right),
\end{eqnarray}
\cor{}{where $\perp$ refers to the orientation of the condensate in the spin space very much like
in Sec.~\ref{Sec:1color}.} The NG mode $\nu$ and the massive mode $\rho$ are coupled and give the
following spectrum, \begin{eqnarray}
E_{\rho}^{2} & = & \left(\frac{m_{2}^{2}}{c_{2}}+\frac{c_{1}^{2}}{2c_{2}^{2}}\right)
+\left(1+\frac{c_{1}^{2}}{c_{1}^{2}+2c_{2}m_{2}^{2}}\right)
\left(\frac{a_{1}+a_{2}}{c_{2}}k_{\perp}^{2}+\frac{a_{1}}{c_{2}}k_{3}^{2}\right),\nonumber \\
E_{\nu}^{2} & = &
\frac{2c_{2}m_{2}^{2}}{2c_{2}m_{2}^{2}+c_{1}^{2}}\left(\frac{a_{1}+a_{2}}{c_{2}}
k_{\perp}^{2}+\frac{a_{1}}{c_{2}}k_{3}^{2}\right),
\end{eqnarray}
where we have defined $m_{1}^{2}=\frac{b(d_{2}+d_{3})}{\bar{d}}$,
$m_{2}^{2}=\frac{2bd_{3}}{\bar{d}}$. The NG modes $\kappa_{a}$ give
\begin{eqnarray}
E_{\kappa'}^{2} & = & \frac{c_{1}^{2}}{c_{2}^{2}}+2\left(\frac{a_{1}}{c_{2}}k_{\perp}^{2}
+\frac{a_{1}+a_{2}}{c_{2}}k_{3}^{2}\right),\nonumber \\
E_{\kappa}^{2} & = &
\left(\frac{a_{1}}{c_{1}}\right)^{2}k_{\perp}^{4}
+\left(\frac{a_{1}+a_{2}}{c_{1}}\right)^{2}k_{3}^{4}.
\end{eqnarray}
The massive modes $\chi_{ij}$ and $\varphi_{ij}$ are coupled to each
other and give the spectrum
\begin{eqnarray*} E_{\chi,\varphi}^{2} & = &
\left[\frac{m_{1}^{2}}{c_{2}}\pm\frac{c_{1}^{2}}{2c_{2}^{2}}\left(
\sqrt{1+\frac{4c_{2}m_{1}^{2}}{c_{1}^{2}}}\pm1\right)\right]
+\left(1\pm\sqrt{\frac{c_{1}^{2}}{c_{1}^{2}+4c_{2}m_{1}^{2}}}\right)
\left(\frac{a_{1}+a_{2}}{c_{2}}k_{\perp}^{2}+\frac{a_{1}}{c_{2}}k_{3}^{2}\right).
\end{eqnarray*}


\subsection{A-phase}
\label{Subapp:disp_A} The order parameter can be written in the
following form \begin{eqnarray} \bm{\Delta} & = &
\exp(i\theta)\exp\left(\frac{1}{2}i\kappa_{a}\lambda_{a}\right)
(v\bm{\Delta}_{\mathrm{A}}+H)\exp(i\nu_{i}J_{i}),\end{eqnarray}
where the summations over $a$ and $i$ are in the ranges $a=4,5,6,7$
and $i=1,2$. The matrix $H$ for massive fields can be parameterized
as
\begin{eqnarray}
H & = & \left(\begin{array}{ccc}
\varphi_{4}-i\varphi_{5} & -i(\varphi_{4}-i\varphi_{5}) & \chi_{11}+i\chi_{12}\\
\varphi_{6}-i\varphi_{7} & -i(\varphi_{6}-i\varphi_{7}) & \chi_{21}+i\chi_{22}\\
h+\rho_{1}+i\rho_{2} & ih-i(\rho_{1}+i\rho_{2}) &
0\end{array}\right).
\end{eqnarray}
The NG fields $\kappa_{4},\kappa_{5}$ and the massive fields
$\varphi_{4},\varphi_{5}$ are coupled (in exactly the same way the
NG fields $\kappa_{6},\kappa_{7}$ and the massive fields
$\varphi_{6},\varphi_{7}$, as they form doublets of the unbroken $\gr{SU(2)_L}$) and give the
dispersion relations,
\begin{eqnarray}
E_{\kappa'}^{2} & = &
\frac{c_{1}^{2}}{c_{2}^{2}}+2\left(\frac{a_{1}+a_{2}}{c_{2}}k_{\perp}^{2}
+\frac{a_{1}}{c_{2}}k_{3}^{2}\right),\nonumber \\
E_{\kappa}^{2} & = &
\left(\frac{a_{1}+a_{2}}{c_{1}}\right)^{2}k_{\perp}^{4}
+\left(\frac{a_{1}}{c_{1}}\right)^{2}k_{3}^{4},\nonumber \\
E_{\varphi}^{2} & = &
\frac{1}{2}\left[\frac{m_{2}^{2}}{c_{2}}\pm\frac{c_{1}^{2}}{c_{2}^{2}}
\left(\sqrt{1+\frac{2c_{2}m_{2}^{2}}{c_{1}^{2}}}\pm1\right)\right]
+\left(1\pm\sqrt{\frac{c_{1}^{2}}{c_{1}^{2}+2c_{2}m_{2}^{2}}}\right)
\left(\frac{a_{1}+a_{2}}{c_{2}}k_{\perp}^{2}+\frac{a_{1}}{c_{2}}k_{3}^{2}\right),
\end{eqnarray}
where we have used $m_{1}^{2}=\frac{bd_{2}}{\bar{d}}$,
$m_{2}^{2}=\frac{2b(d_{2}-d_{3})}{\bar{d}}$,
$m_{3}^{2}=-\frac{4bd_{3}}{\bar{d}}$. The NG mode $\theta$ and the
massive modes $h,\rho$ are coupled and give the eigenmodes,
\begin{eqnarray*}
E_{\theta}^{2} & = &
\left(1+\frac{c_{1}^{2}}{2bc_{2}-c_{1}^{2}}\right)
\left(\frac{a_{1}+a_{2}}{c_{2}}k_{\perp}^{2}+\frac{a_{1}}{c_{2}}k_{3}^{2}\right),\\
E_{h}^{2} & = &
\frac{-2b}{c_{2}}+\frac{c_{1}^{2}}{c_{2}^{2}}+\left(1-\frac{c_{1}^{2}}{2bc_{2}-c_{1}^{2}}\right)
\left(\frac{a_{1}+a_{2}}{c_{2}}k_{\perp}^{2}+\frac{a_{1}}{c_{2}}k_{3}^{2}\right),\\
E_{\rho}^{2} & = &
\frac{m_{3}^{2}}{2c_{2}}+\frac{c_{1}^{2}}{2c_{2}^{2}}\left[1\pm\sqrt{1+\frac{(c_{2}m_{3}^{2}+4bc_{2}
)^{2}+4c_{1}^{2}c_{2}m_{3}^{2}}{4c_{1}^{4}}}\right]\\
&  & +\left[1\pm\sqrt{\frac{4c_{1}^{4}}{(c_{2}m_{3}^{2}+4bc_{2})^{2}
+4c_{1}^{2}c_{2}m_{3}^{2}+4c_{1}^{4}}}\right]
\left(\frac{a_{1}+a_{2}}{c_{2}}k_{\perp}^{2}+\frac{a_{1}}{c_{2}}k_{3}^{2}\right).
\end{eqnarray*}
The NG modes $\nu_{1,2}$ give \begin{eqnarray} E_{\nu'}^{2} & = &
\frac{c_{1}^{2}}{c_{2}^{2}}+2\left(\frac{a_{1}}{c_{2}}k_{\perp}^{2}
+\frac{a_{1}+a_{2}}{c_{2}}k_{3}^{2}\right),\nonumber \\
E_{\nu}^{2} & = &
\left(\frac{a_{1}}{c_{1}}\right)^{2}k_{\perp}^{4}
+\left(\frac{a_{1}+a_{2}}{c_{1}}\right)^{2}k_{3}^{4}.
\end{eqnarray}
The massive modes $\chi_{ij}$ $(i,j=1,2)$ give
\begin{eqnarray*}
E_{\chi}^{2} & = & \left[\frac{m_{1}^{2}}{c_{2}}\pm\frac{c_{1}^{2}}{2c_{2}^{2}}
\left(\sqrt{1+\frac{4c_{2}m_{1}^{2}}{c_{1}^{2}}}\pm1\right)\right]
+\left(1\pm\sqrt{\frac{c_{1}^{2}}{c_{1}^{2}+4c_{2}m_{1}^{2}}}\right)
\left(\frac{a_{1}}{c_{2}}k_{\perp}^{2}+\frac{a_{1}+a_{2}}{c_{2}}k_{3}^{2}\right).
\end{eqnarray*}


\subsection{$\eps$-phase}
\label{Subapp:disp_eps} The order parameter can be written in the
following form
\begin{eqnarray} \bm{\Delta} & = &
\exp(i\theta_{2}\mathcal{P}_{2})\exp(i\theta_{3}\mathcal{P}_{3})
\exp\left(i\frac{1}{2}\kappa_{a}\lambda_{a}\right)(v\bm{\Delta}_{\varepsilon}+H)\exp(i\nu_{i}J_{i})
\end{eqnarray}
where the summations over $a$ and $i$ are in the ranges
$a=1,2,4,5,6,7$ and $i=1,2$. The matrix $H$ for massive fields can
be parameterized as
\begin{eqnarray}
H & = & \left(\begin{array}{ccc}
\varphi_{4}-i\varphi_{5} & -i(\varphi_{4}-i\varphi_{5}) & 0\\
\varphi_{6}-i\varphi_{7} & -i(\varphi_{6}-i\varphi_{7}) & h_{2}\\
h_{3}+\rho_{1}+i\rho_{2} & ih_{3}-i(\rho_{1}+i\rho_{2}) &
0\end{array}\right).
\end{eqnarray}
We can define masses
\begin{eqnarray*}
m_{2}^{2} & = &
-\frac{2b[d_{1}d_{2}+d_{2}(d_{2}+d_{3})]}{\bar{d}(2d_{2}+d_{3})},\;
m_{3}^{2}=-\frac{4b(d_{1}+d_{2})(d_{2}+d_{3})}{\bar{d}(2d_{2}+d_{3})},\\
m_{23}^{2} & = &
-\frac{4\sqrt{2}bd_{1}\sqrt{d_{2}(d_{2}+d_{3})}}{\bar{d}(2d_{2}+d_{3})},\;
m_{45}^{2}=-\frac{2b(d_{3}^{2}-d_{2}^{2})}{\bar{d}(2d_{2}+d_{3})},\\
m_{67}^{2} & = & -\frac{2bd_{3}^{2}}{\bar{d}(2d_{2}+d_{3})},\;
m_{\rho}^{2} =-\frac{4bd_{2}(d_{2}+d_{3})}{\bar{d}(2d_{2}+d_{3})}.
\end{eqnarray*}
The NG modes $\theta_{2},\theta_{3}$ and the massive modes
$h_{2},h_{3},\rho_{1},\rho_{2}$ are coupled and give the following
eigenmodes,
\begin{eqnarray*}
E_{\theta_{2}}^{2} & = & \frac{m_{23}^{2}+4m_{2}^{2}}{m_{23}^{2}
+4m_{2}^{2}+4c_{1}^{2}/c_{2}^{2}}\left(\frac{a_{1}}{c_{2}}k_{\perp}^{2}+\frac{a_{1}+a_{2}}{c_{2}}
k_{3}^{2}\right),\\
E_{h}^{2} & = & \frac{m_{23}^{2}+4m_{2}^{2}}{4c_{2}}+\frac{c_{1}^{2}}{c_{2}^{2}}
+\left(1+\frac{4c_{1}^{2}/c_{2}^{2}}{m_{23}^{2}+4m_{2}^{2}+4c_{1}^{2}/c_{2}^{2}}\right)
\left(\frac{a_{1}}{c_{2}}k_{\perp}^{2}+\frac{a_{1}+a_{2}}{c_{2}}k_{3}^{2}\right),\\
E_{\theta_{3}}^{2} & = &
\frac{m_{3}^{2}+m_{23}^{2}}{m_{3}^{2}+m_{23}^{2}+2c_{1}^{2}/c_{2}}
\left(\frac{a_{1}+a_{2}}{c_{2}}k_{\perp}^{2}+\frac{a_{1}}{c_{2}}k_{3}^{2}\right),\\
E_{h}^{2} & = &
\frac{m_{3}^{2}+m_{23}^{2}}{4c_{2}}+\frac{c_{1}^{2}}{c_{2}^{2}}+\left(1+\frac{2c_{1}^{2}/c_{2}
}{m_{3}^{2}+m_{23}^{2}+2c_{1}^2/c_{2}}\right)\left(\frac{a_{1}+a_{2}}{c_{2}}
k_{\perp}^{2}+\frac{a_{1}}{c_{2}}k_{3}^{2}\right),\\
E_{\rho}^{2} & = &
\frac{m_{\rho}^{2}}{2c_{2}}+\frac{c_{1}^{2}}{2c_{2}^{2}}\left(1\pm\sqrt{1+\frac{2c_{2}m_{\rho}^{2}}{
c_{1}^{2}}}\right)
+\left(1\pm\sqrt{\frac{c_{1}^{2}}{c_{1}^{2}+2c_{2}m_{\rho}^{2}}}\right)
\left(\frac{a_{1}+a_{2}}{c_{2}}k_{\perp}^{2}+\frac{a_{1}}{c_{2}}k_{3}^{2}\right).
\end{eqnarray*}
For NG bosons $\kappa_{1,2}$ we obtain the energies,
\begin{eqnarray}
E_{\kappa}^{2} & = &
\frac{c_{1}^{2}}{c_{2}^{2}}+2\left(\frac{a_{1}}{c_{2}}k_{\perp}^{2}
+\frac{a_{1}+a_{2}}{c_{2}}k_{3}^{2}\right),\nonumber \\
E_{\kappa'}^{2} & = &
\left(\frac{a_{1}}{c_{1}}\right)^{2}k_{\perp}^{4}
+\left(\frac{a_{1}+a_{2}}{c_{1}}\right)^{2}k_{3}^{4}.
\end{eqnarray}
The NG modes $\kappa_{4,5}$ and the massive modes $\varphi_{4,5}$
are coupled (in the same way as $\kappa_{6,7}$ are coupled with
$\varphi_{6,7}$) and give the dispersion relations
\begin{eqnarray}
E_{\kappa}^{2} & = &
\frac{c_{1}^{2}}{c_{2}^{2}}+2\left(\frac{a_{1}+a_{2}}{c_{2}}k_{\perp}^{2}
+\frac{a_{1}}{c_{2}}k_{3}^{2}\right),\nonumber \\
E_{\kappa'}^{2} & = &
\left(\frac{a_{1}+a_{2}}{c_{1}}\right)^{2}
k_{\perp}^{4}+\left(\frac{a_{1}}{c_{1}}\right)^{2}k_{3}^{4},\nonumber \\
E_{\varphi}^{2} & = &
\frac{m_{45}^{2}}{c_{2}}+\frac{c_{1}^{2}}{2c_{2}^{2}}\left(1\pm\sqrt{1+\frac{2c_{2}m_{45}^{2}}{c_{1}
^ {2}}}\right)
+1\pm\sqrt{\frac{c_{1}^{2}}{c_{1}^{2}+2c_{2}m_{45}^{2}}}.
\end{eqnarray}
The NG modes $\nu_{1,2}$ give
\begin{eqnarray}
E_{\nu'}^{2} & = &
\frac{c_{1}^{2}}{c_{2}^{2}}+2\left(\frac{a_{1}}{c_{2}}k_{\perp}^{2}+\frac{a_{1}+a_{2}}{c_{2}}k_{3}^{
2}\right),\nonumber \\
E_{\nu}^{2} & = &
\left(\frac{a_{1}}{c_{1}}\right)^{2}k_{\perp}^{4}
+\left(\frac{a_{1}+a_{2}}{c_{1}}\right)^{2}k_{3}^{4}.
\end{eqnarray}


\bibliography{helix}

\begin{thebibliography}{10}
\expandafter\ifx\csname url\endcsname\relax
  \def\url#1{\texttt{#1}}\fi
\expandafter\ifx\csname urlprefix\endcsname\relax\def\urlprefix{URL }\fi
\expandafter\ifx\csname href\endcsname\relax
  \def\href#1#2{#2} \def\path#1{#1}\fi

\bibitem{Alford:2007xm}
M.~G. Alford, A.~Schmitt, K.~Rajagopal, T.~{Sch\"afer}, {Color
  superconductivity in dense quark matter}, Rev. Mod. Phys. 80 (2008)
  1455--1515.
\newblock \href {http://arxiv.org/abs/0709.4635} {\path{arXiv:0709.4635}}.

\bibitem{Wang:2009xf}
Q.~Wang, {Some aspects of color superconductivity: an introduction}, Prog.
  Phys. 30 (2010) 173--216.
\newblock \href {http://arxiv.org/abs/0912.2485} {\path{arXiv:0912.2485}}.

\bibitem{Huang:2010nn}
M.~Huang, {QCD phase diagram at high temperature and density}, Prog. Phys. 30
  (2010) 217--246.
\newblock \href {http://arxiv.org/abs/1001.3216} {\path{arXiv:1001.3216}}.

\bibitem{Fukushima:2010bq}
K.~Fukushima, T.~Hatsuda, {The phase diagram of dense QCD}\href
  {http://arxiv.org/abs/1005.4814} {\path{arXiv:1005.4814}}.

\bibitem{Rajagopal:2005dg}
K.~Rajagopal, A.~Schmitt, {Stressed pairing in conventional color
  superconductors is unavoidable}, Phys. Rev. D73 (2006) 045003.
\newblock \href {http://arxiv.org/abs/hep-ph/0512043}
  {\path{arXiv:hep-ph/0512043}}.

\bibitem{Feng:2007bg}
B.~Feng, D.-f. Hou, H.-c. Ren, {Angular momentum mixing in a non-spherical
  color superconductor}, Nucl. Phys. B796 (2008) 500--520.
\newblock \href {http://arxiv.org/abs/0711.0496} {\path{arXiv:0711.0496}}.

\bibitem{Bailin:1983bm}
D.~Bailin, A.~Love, {Superfluidity and superconductivity in relativistic
  fermion systems}, Phys. Rept. 107 (1984) 325--385.

\bibitem{Schaefer:2000tw}
T.~{Sch\"afer}, {Quark hadron continuity in QCD with one flavor}, Phys. Rev.
  D62 (2000) 094007.
\newblock \href {http://arxiv.org/abs/hep-ph/0006034}
  {\path{arXiv:hep-ph/0006034}}.

\bibitem{Hosek:2000fn}
J.~{Ho\v{s}ek}, {Anisotropic color superconductor}\href
  {http://arxiv.org/abs/hep-ph/0011034} {\path{arXiv:hep-ph/0011034}}.

\bibitem{Alford:2002rz}
M.~G. Alford, J.~A. Bowers, J.~M. Cheyne, G.~A. Cowan, {Single color and single
  flavor color superconductivity}, Phys. Rev. D67 (2003) 054018.
\newblock \href {http://arxiv.org/abs/hep-ph/0210106}
  {\path{arXiv:hep-ph/0210106}}.

\bibitem{Buballa:2002wy}
M.~Buballa, J.~{Ho\v{s}ek}, M.~Oertel, {Anisotropic Admixture in
  Color-Superconducting Quark Matter}, Phys. Rev. Lett. 90 (2003) 182002.
\newblock \href {http://arxiv.org/abs/hep-ph/0204275}
  {\path{arXiv:hep-ph/0204275}}.

\bibitem{Alford:2005yy}
M.~G. Alford, G.~A. Cowan, {Single-flavour and two-flavour pairing in
  three-flavour quark matter}, J. Phys. G32 (2006) 511--528.
\newblock \href {http://arxiv.org/abs/hep-ph/0512104}
  {\path{arXiv:hep-ph/0512104}}.

\bibitem{Schmitt:2004et}
A.~Schmitt, {The ground state in a spin-one color superconductor}, Phys. Rev.
  D71 (2005) 054016.
\newblock \href {http://arxiv.org/abs/nucl-th/0412033}
  {\path{arXiv:nucl-th/0412033}}.

\bibitem{Brauner:2008ma}
T.~Brauner, {Helical ordering in the ground state of spin-one color
  superconductors as a consequence of parity violation}, Phys. Rev. D78 (2008)
  125027.
\newblock \href {http://arxiv.org/abs/0810.3481} {\path{arXiv:0810.3481}}.

\bibitem{Schmitt:2002sc}
A.~Schmitt, Q.~Wang, D.~H. Rischke, {When the transition temperature in color
  superconductors is not like in BCS theory}, Phys. Rev. D66 (2002) 114010.
\newblock \href {http://arxiv.org/abs/nucl-th/0209050}
  {\path{arXiv:nucl-th/0209050}}.

\bibitem{Schmitt:2003xq}
A.~Schmitt, Q.~Wang, D.~H. Rischke, {Electromagnetic Meissner Effect in
  Spin-One Color Superconductors}, Phys. Rev. Lett. 91 (2003) 242301.
\newblock \href {http://arxiv.org/abs/nucl-th/0301090}
  {\path{arXiv:nucl-th/0301090}}.

\bibitem{Schmitt:2003aa}
A.~Schmitt, Q.~Wang, D.~H. Rischke, {Mixing and screening of photons and gluons
  in a color superconductor}, Phys. Rev. D69 (2004) 094017.
\newblock \href {http://arxiv.org/abs/nucl-th/0311006}
  {\path{arXiv:nucl-th/0311006}}.

\bibitem{Schmitt:2005ee}
A.~Schmitt, I.~A. Shovkovy, Q.~Wang, {Pulsar Kicks Via Spin-1 Color
  Superconductivity}, Phys. Rev. Lett. 94 (2005) 211101.
\newblock \href {http://arxiv.org/abs/hep-ph/0502166}
  {\path{arXiv:hep-ph/0502166}}.

\bibitem{Schmitt:2005wg}
A.~Schmitt, I.~A. Shovkovy, Q.~Wang, {Neutrino emission and cooling rates of
  spin-one color superconductors}, Phys. Rev. D73 (2006) 034012.
\newblock \href {http://arxiv.org/abs/hep-ph/0510347}
  {\path{arXiv:hep-ph/0510347}}.

\bibitem{Aguilera:2005tg}
D.~N. Aguilera, D.~Blaschke, M.~Buballa, V.~L. Yudichev, {Color-spin locking
  phase in two-flavor quark matter for compact star phenomenology}, Phys. Rev.
  D72 (2005) 034008.
\newblock \href {http://arxiv.org/abs/hep-ph/0503288}
  {\path{arXiv:hep-ph/0503288}}.

\bibitem{Aguilera:2006cj}
D.~N. Aguilera, D.~Blaschke, H.~Grigorian, N.~N. Scoccola, {Nonlocality effects
  on color spin locking condensates}, Phys. Rev. D74 (2006) 114005.
\newblock \href {http://arxiv.org/abs/hep-ph/0604196}
  {\path{arXiv:hep-ph/0604196}}.

\bibitem{Aguilera:2006xv}
D.~N. Aguilera, {Spin-one color superconductivity in compact stars? An analysis
  within NJL-type models}, Astrophys. Space Sci. 308 (2007) 443--450.
\newblock \href {http://arxiv.org/abs/hep-ph/0608041}
  {\path{arXiv:hep-ph/0608041}}.

\bibitem{Sa'd:2006qv}
B.~A. Sa'd, I.~A. Shovkovy, D.~H. Rischke, {Bulk viscosity of spin-one color
  superconductors with two quark flavors}, Phys. Rev. D75 (2007) 065016.
\newblock \href {http://arxiv.org/abs/astro-ph/0607643}
  {\path{arXiv:astro-ph/0607643}}.

\bibitem{Wang:2006tg}
Q.~Wang, Z.-g. Wang, J.~Wu, {Phase space and quark mass effects in neutrino
  emissions in a color superconductor}, Phys. Rev. D74 (2006) 014021.
\newblock \href {http://arxiv.org/abs/hep-ph/0605092}
  {\path{arXiv:hep-ph/0605092}}.

\bibitem{Blaschke:2008br}
D.~Blaschke, F.~Sandin, T.~{Kl\"ahn}, J.~Berdermann, {Sequential deconfinement
  of quark flavors in neutron stars}, Phys. Rev. C80 (2009) 065807.
\newblock \href {http://arxiv.org/abs/0807.0414} {\path{arXiv:0807.0414}}.

\bibitem{Marhauser:2006hy}
F.~Marhauser, D.~Nickel, M.~Buballa, J.~Wambach, {Color-spin locking in a
  self-consistent Dyson-Schwinger approach}, Phys. Rev. D75 (2007) 054022.
\newblock \href {http://arxiv.org/abs/hep-ph/0612027}
  {\path{arXiv:hep-ph/0612027}}.

\bibitem{Feng:2009vt}
B.~Feng, D.~Hou, H.-c. Ren, P.-p. Wu, {Single-Flavor Color Superconductivity in
  a Magnetic Field}, Phys. Rev. Lett. 105 (2010) 042001.
\newblock \href {http://arxiv.org/abs/0911.4997} {\path{arXiv:0911.4997}}.

\bibitem{Wang:2009if}
X.~Wang, H.~Malekzadeh, I.~A. Shovkovy, {Non-leptonic weak processes in
  spin-one color superconducting quark matter}, Phys. Rev. D81 (2010) 045021.
\newblock \href {http://arxiv.org/abs/0912.3851} {\path{arXiv:0912.3851}}.

\bibitem{Wang:2010yd}
X.~Wang, I.~A. Shovkovy, {Bulk viscosity of spin-one color superconducting
  strange quark matter}\href {http://arxiv.org/abs/1006.1293}
  {\path{arXiv:1006.1293}}.

\bibitem{Dzyaloshinsky:1958dz}
I.~Dzyaloshinsky, A thermodynamic theory of weak ferromagnetism of
  antiferromagnetics, J. Phys. Chem. Solids 4 (1958) 241--255.

\bibitem{Moriya:1960mo}
T.~Moriya, Anisotropic superexchange interaction and weak ferromagnetism, Phys.
  Rev. 120 (1960) 91--98.

\bibitem{Low:2001bw}
I.~Low, A.~V. Manohar, {Spontaneously Broken Spacetime Symmetries and
  Goldstone's Theorem}, Phys. Rev. Lett. 88 (2002) 101602.
\newblock \href {http://arxiv.org/abs/hep-th/0110285}
  {\path{arXiv:hep-th/0110285}}.

\bibitem{Nielsen:1975hm}
H.~B. Nielsen, S.~Chadha, {On how to count Goldstone bosons}, Nucl. Phys. B105
  (1976) 445--453.

\bibitem{Brauner:2010wm}
T.~Brauner, {Spontaneous Symmetry Breaking and Nambu--Goldstone Bosons in
  Quantum Many-Body Systems}, Symmetry 2 (2010) 609--657.
\newblock \href {http://arxiv.org/abs/1001.5212} {\path{arXiv:1001.5212}}.

\bibitem{Brauner:2009df}
T.~Brauner, J.-y. Pang, Q.~Wang, {Symmetry breaking patterns and collective
  modes of spin-one color superconductors}, Nucl. Phys. A844 (2010) 216c--223c.
\newblock \href {http://arxiv.org/abs/0909.4201} {\path{arXiv:0909.4201}}.

\bibitem{Son:2002zn}
D.~Son, {Low-energy quantum effective action for relativistic superfluids}\href
  {http://arxiv.org/abs/hep-ph/0204199} {\path{arXiv:hep-ph/0204199}}.

\bibitem{Brauner:2005di}
T.~Brauner, {Goldstone boson counting in linear sigma models with chemical
  potential}, Phys. Rev. D72 (2005) 076002.
\newblock \href {http://arxiv.org/abs/hep-ph/0508011}
  {\path{arXiv:hep-ph/0508011}}.

\bibitem{Weinberg:1972fn}
S.~Weinberg, {Approximate Symmetries and Pseudo-Goldstone bosons}, Phys. Rev.
  Lett. 29 (1972) 1698--1701.

\bibitem{Vollhardt:1990vw}
D.~Vollhardt, P.~W{\"o}lfle, The Superfluid Phases of Helium 3, Taylor and
  Francis, 1990.

\bibitem{Manuel:2004iv}
C.~Manuel, A.~Dobado, F.~J. Llanes-Estrada, {Shear viscosity in a CFL quark
  star}, JHEP 09 (2005) 076.
\newblock \href {http://arxiv.org/abs/hep-ph/0406058}
  {\path{arXiv:hep-ph/0406058}}.

\bibitem{Kapusta:1981aa}
J.~I. Kapusta, {Bose-Einstein condensation, spontaneous symmetry breaking, and
  gauge theories}, Phys. Rev. D24 (1981) 426--439.

\bibitem{Gusynin:2003yu}
V.~P. Gusynin, V.~A. Miransky, I.~A. Shovkovy, {Spontaneous rotational symmetry
  breaking and roton like excitations in gauged $\sigma$-model at finite
  density}, Phys. Lett. B581 (2004) 82--92.
\newblock \href {http://arxiv.org/abs/hep-ph/0311025}
  {\path{arXiv:hep-ph/0311025}}.

\end{thebibliography}
\bibliographystyle{elsarticle-num}

\end{document}